\documentclass[12pt]{article}
\usepackage{amscd,amssymb,amsmath,latexsym,enumerate}
\usepackage[mathscr]{euscript}
\usepackage{epsfig}
\usepackage{fancybox}
\usepackage{verbatim}

\usepackage{tikz}
\usepackage{tikz-cd}
\usepackage{todonotes}

\usepackage{color}

\usepackage{xcolor}
\definecolor{MyBlue}{cmyk}{1,0.13,0,0.63}
\definecolor{MyGreen}{cmyk}{0.91,0,0.88,0.52}
\newcommand{\mylinkcolor}{MyBlue}
\newcommand{\mycitecolor}{MyGreen}
\newcommand{\myurlcolor}{black}

\usepackage{hyperref}
\hypersetup{%
  bookmarksnumbered=true,bookmarksopen=false,%
  plainpages=false,
  linktocpage=true,%
  colorlinks=true,breaklinks=true,%
  linkcolor=\mylinkcolor,citecolor=\mycitecolor,urlcolor=\myurlcolor,%
  pdfpagelayout=OneColumn,%
  pageanchor=true,%
}

\title{The spectral localizer for semifinite spectral triples}

\author{Hermann Schulz-Baldes and Tom Stoiber
\\
\\
{\small Department Mathematik, Friedrich-Alexander-Universit\"at Erlangen-N\"urnberg, Germany}
}

\date{ }

\newtheorem{theorem}{Theorem}
\newtheorem{definition}[theorem]{Definition}
\newtheorem{proposition}[theorem]{Proposition}
\newtheorem{lemma}[theorem]{Lemma}

\newcommand{\CM}{{\mathbb C}}

\newcommand{\ZM}{{\mathbb Z}}
\newcommand{\PM}{{\mathbb P}}

\newcommand{\Aa}{{\cal A}}

\newcommand{\Bb}{{\cal B}}

\newcommand{\Nn}{{\cal N}}

\newcommand{\Kk}{{\cal K}}
\newcommand{\Hh}{{\cal H}}

\newcommand{\one}{{\bf 1}}

\newcommand{\Tr}{\mbox{\rm Tr}}
\newcommand{\Tt}{\tau} 

\newcommand{\SF}{{\rm Sf}}

\newcommand{\Ind}{{\rm Ind}} 
\newcommand{\Ker}{{\rm Ker}} 
\newcommand{\Supp}{{\rm Supp}} 
\newcommand{\Ran}{{\rm Ran}} 
\newcommand{\sgn}{{\rm sgn}} 
\newcommand{\Sig}{{\rm Sig}} 
\newcommand{\ec}{{\rm ec}}
\newcommand{\Fsa}{\mathcal{F}^\textup{sa}}
\newcommand{\tInd}{\textup{Ind}}
\newcommand{\diag}{{\rm diag}}

\newcommand{\Aat}{{\mathcal A}^\sim}
 
\newcommand{\unboundedtrans}{f}

\newcommand{\bbC}{\mathbb{C}}

\newcommand{\bbR}{\mathbb{R}}

\newcommand{\bbZ}{\mathbb{Z}}

\newcommand{\calA}{\mathcal{A}}

\newcommand{\calF}{\mathcal{F}}
\newcommand{\calK}{\mathcal{K}}

\newcommand{\difd}{\textup{d}}

\providecommand{\abs}[1]{\left \lvert#1 \right \rvert} 
\providecommand{\norm}[1]{\left \lVert#1 \right \rVert}

\begin{document}

\maketitle

\begin{abstract}
The notion of spectral localizer is extended to pairings with semifinite spectral triples. By a spectral flow argument, any semifinite index pairing is shown to be equal to the signature of the spectral localizer. As an application, a formula for the weak invariants of topological insulators is derived. This provides a new approach to their numerical evaluation.
\hfill MSC2010: 19K56, 46L80

\end{abstract}

\section{Overview}
\label{sec-Overview}

In two recent works \cite{LS,LS2}, it was shown that the integer-valued pairing of a $K$-theory class with a Fredholm module is equal to the signature of a certain gapped self-adjoint operator called the spectral localizer. It is the object of this work to generalize this equality to semifinite index theory. Let us describe the result on integer-valued pairings more explicitly in the case of an odd index pairing. Hence let $a$ be an invertible operator on a Hilbert space and $D=D^*$ an unbounded selfadjoint operator with compact resolvent and bounded commutator $[D,a]$. If $P=\chi(D>0)$ is the positive spectral projection of $D$ expressed in terms of the characteristic function $\chi$, then $PaP+(1-P)$ is a Fredholm operator and the index pairing is $\langle [a],[D]\rangle=\Ind(PaP+(1-P))$. The main result of \cite{LS,LS3} then states that this pairing can be computed with the spectral localizer
$$
L_\kappa
\;=\;
\begin{pmatrix}
\kappa\,D & a \\ a^* & -\kappa\,D
\end{pmatrix}
\;,
$$
namely if $L_{\kappa,\rho}$ is the finite-dimensional restriction of $L_\kappa$ to the range of $\chi(|D\oplus D|<\rho)$, then provided $\rho$ is sufficiently large and $\kappa$ sufficiently small,
\begin{equation}
\label{eq-IntroFormula}
\langle [a],[D]\rangle
\;=\;
\tfrac{1}{2}\,\Sig(L_{\kappa,\rho})
\;.
\end{equation}
The proof of \eqref{eq-IntroFormula} contains an analytic argument showing that $0$ is not in the spectrum of $L_{\kappa,\rho}$ so that the signature is well-defined, and then a topological homotopy argument leading to the equality  \eqref{eq-IntroFormula}. In the works \cite{LS,LS2}, this latter step was based on an evaluation of the index map and actually \eqref{eq-IntroFormula} then appeared merely as corollary to a $K$-theoretic statement. An alternative argument for \eqref{eq-IntroFormula} using spectral flow was provided in \cite{LS3,LSS}. The main interest of \eqref{eq-IntroFormula} is that the r.h.s. is amenable to numerical evaluation and is of use in connection with topological insulators, see \cite{Lor,LSS}. In fact, in applications the finite volume restriction of $L_\kappa$ is readily obtained and no further functional calculus is involved so that merely the signature of a finite-dimensional matrix has to computed, either by diagonalizing $L_{\kappa,\rho}$ or by the block Cholesky decomposition. 

\vspace{.2cm}

For the generalization of \eqref{eq-IntroFormula}, Section~\ref{sec-sfgeneral} first reviews the definitions and main results about Fredholm operators and spectral flow in semifinite von Neumann algebras, based mainly on \cite{PR,Phi2,BCP,CPRS,KNR}. It then introduces the semifinite signature and analyses its link to spectral flow. With this arsenal established, the main results can be stated in Section~\ref{sec-SpecLoc}. Section~\ref{sec-Constancy} adapts the arguments of \cite{LS2} to prove that the signature is well-defined and stable. Section~\ref{sec-ProofOdd} then generalizes the spectral flow proof of \eqref{eq-IntroFormula} given in \cite{LS3} to the present setting, while Section~\ref{sec-ProofEven} provides the proof that also the even index pairings can be calculated with the half-signature. The spectral flow argument in Section~\ref{sec-ProofEven} is different from the one given in \cite{LSS}, and we feel that it is more direct. Moreover, it does not require the normality of the off-diagonal entries of the (even) Dirac operator as in \cite{LS2,LSS}, but instead supposes that the spectral triple is Lipschitz-regular in the notation of \cite{CM}. Finally Section~\ref{sec-app} discusses applications to the numerical computation of weak invariants of topological insulators.

\section{Semifinite spectral flow and signature}
\label{sec-sfgeneral}

Let $\Nn$ be a von Neumann algebra with a semifinite faithful normal trace $\Tt$. A projection $P \in \Nn$ is called finite if $\tau(P)<\infty$.  Let $\Kk$ denote the norm-closure of the smallest algebraic ideal in $\Nn$ containing the finite projections. This C$^*$-algebra is called the ideal of $\tau$-compact operators in $\Nn$ and any projection in $\Kk$ is finite. Associated to $\Kk$ is a short exact sequence $0\to\Kk\to\Nn\to\Nn/\Kk\to0$. The quotient $\Nn/\Kk$ is called the Calkin algebra and the quotient map is denoted by $\pi$. Associated to this exact sequence one has the standard notion of Breuer-Fredholm operator, but for the definition of spectral flow an extension to skew-corners is needed and will be described next. Associated to projections $P,Q \in \Nn$, the skew-corner $P\Nn Q$ consists of operators in $\Nn$ mapping $\Ran(Q)$ to $\Ran(P)$ and then $T\in P\Nn Q$ will be viewed as an operator between these subspaces. The most basic example is $PQ\in P\Nn Q$. Note that $P\Nn Q$ is an algebra only if $P=Q$. For any $T \in \Nn$ we will denote by $\Ker(T)$ both the kernel as a subspace as well as the projection onto it, which is an element in $\Nn$. Furthermore, for any pair of projections $P,Q\in\Nn$, let $P\cap Q$ denote the projection onto the intersection of $\Ran(P)$ and $\Ran(Q)$.

\begin{definition}[\cite{CPRS,BCP}]
Let $P,Q \in \Nn$ be projections and $T \in \Nn $. Then $T$ is called $(P\cdot Q)$-Fredholm if $\Ker(T) \cap Q$ and $\Ker(T^*)\cap P$ are $\Tt$-finite projections and there exists a projection $E\in\Nn$ such that $P-E$ is finite and $\Ran(E)\subset\Ran(T)$.
Its (semifinite) index is then defined as
$$
\tInd_{(P\cdot Q)}(T) \;=\; \Tt\big(\Ker(T) \cap Q\big) \,-\, \Tt\big(\Ker(T^*)\cap P\big)
\;.
$$
\end{definition}

If $P=Q=\one$, this is reduces to the standard semifinite index, which is sometimes denoted by $\Tt$-Ind. There exists the following generalization of Atkinson's theorem.

\begin{theorem}[\cite{CPRS}] 
Let $P,Q,R \in \Nn$ be projections and $T \in P\Nn Q $. 

\begin{itemize}

\item[{\rm (i)}] $T$ is $(P\cdot Q)$-Fredholm if and only if there exists $S\in Q\Nn P$ with $TS-P\in P\Kk P$ and $ST-Q\in Q\Kk Q$.

\item[{\rm (ii)}] The set of $(P\cdot Q)$-Fredholm operators is open in $P\Nn Q$.

\item[{\rm (iii)}] If $T$ is $(P\cdot Q)$-Fredholm and $S\in Q\Nn R $ is $(Q \cdot R)$-Fredholm, then $TS$ is $(P \cdot R)$-Fredholm 
$$
\tInd_{(P\cdot R)}(TS) \;=\; \tInd_{(P\cdot Q)}(T) \,+\, \tInd_{(Q\cdot R)}(S)
\;.
$$

\end{itemize}

\end{theorem}

The following criterion is crucial for the definition of the spectral flow:

\begin{proposition}[\cite{Phi2,BCP}]
If $P,Q\in\Nn$ are projections with $\norm{\pi(P-Q)}<1$, then $PQ$ is $(P\cdot Q)$-Fredholm. 
\end{proposition}

\begin{definition}
For projections $P,Q\in\Nn$ with $\norm{\pi(P-Q)}<1$, the essential codimension is
$$
\ec (P,Q)\;:=\;\Ind_{(P,Q)}(PQ)
\;=\;
\Tt\big((1-P)\cap Q\big)\,-\,\Tt\big((1-Q)\cap P\big)
\;.
$$
\end{definition}

Denote by $\Fsa\subset\Nn$ the space of self-adjoint Fredholm operators.

\begin{definition}
\label{def-SF}
Let $\{T_t\}_{t\in [0,1]}$ be a norm-continuous path in $\Fsa$ and  $0=t_0 < t_1 < ... < t_{n+1} = 1$ be a partition such that 
$$
\norm{\pi\left(p_{k+1}-p_{k}\right)} 
\;\leq\; 
\tfrac{1}{2}
\;,
$$ 
for all $k=0,\ldots,n$ with $p_k := \chi(T_{t_k} \geq 0)$. Then the spectral flow is defined as the real number
$$
\SF(\{T_t\}_{t\in [0,1]}) 
\;:= \;
\sum_{k=0}^{n} \ec(p_k,p_{k+1})
\;.
$$
If it is clear from the context, the index $t\in [0,1]$ is dropped.
\end{definition}

Let us quote the following basic properties of the spectral flow:

\begin{proposition}[\cite{BCP}]
\label{prop-SFprop}
Let  $\{T_t\}$ and $\{T'_t\}$ be norm-continuous paths in $\Fsa$.
\begin{itemize}
\item[{\rm (i)}] The spectral flow is well defined and does not depend on the choice of partition.
\item[{\rm (ii)}] {\rm (Homotopy invariance)} If $\{T_t\}$ and $\{T'_t\}$  have the same endpoints and are connected by a norm-continuous homotopy within $\Fsa$, then 
$$
\SF(\{T_t\}) \;=\; \SF(\{T'_t\})
\;.
$$
\item[{\rm (iii)}] {\rm (Concatenation)} If $T_1 = T'_0$, then 
$$
\SF(\{T_{t}\} * \{T'_{t}\}) \;=\; \SF(\{T_{t}\})+\SF(\{T'_{t}\})
\;,
$$
with $*$ denoting concatenation of paths.
\item[{\rm (iv)}] {\rm (Homomorphism)}  
$$
\SF(\{T_t \oplus T'_t\}) \;=\; \SF(\{T_{t}\})+\SF(\{T'_{t}\})
\;.
$$
\end{itemize}
\end{proposition}

If one has a path $\{D_t\}$ of self-adjoint unbounded operators affiliated to $\Nn$ (notably, each spectral projection of $D_t$ lies in $\Nn$) such that  its bounded transform
\begin{equation}
\label{eq-unboundedtrans}
\unboundedtrans(D_t)
\; := \; 
D_t (1+D_t^2)^{-1/2}
\end{equation}
is a norm-continuous path in $\Fsa$, then its spectral flow can be defined by 
$$
\SF(\{D_t\})
\;:=\; 
\SF(\{f(D_t)\})
\;.
$$
It then satisfies all the properties of Proposition~\ref{prop-SFprop}. In this work it is enough to consider bounded perturbations of a fixed unbounded operator (a more general definition is given in \cite{BCP}):

\begin{proposition}
\label{eq-DiracPerturb}
Let $D$ be an invertible densely defined self-adjoint operator affiliated to $\Nn$. 
\begin{enumerate}[{\rm (i)}]
\item For $D_t := D +a_t$ with $\{a_t\}_{t\in [0,1]}$ a norm-continuous path of self-adjoint elements in $\Nn$ the bounded transform $\{f(D_t)\}_{t\in [0,1]}$ is a norm-continuous path in $\Nn$.
\item If $a \in \Nn$ satisfies $a (1+D^2)^{-\frac{1}{2}} \in \calK$, then $f(D+a)$ is Fredholm.
\end{enumerate}
\end{proposition}

\noindent{\bf Proof.}
By \cite[Lemma 2.7]{CP} the difference $f(D + a) - f(D)$ is given as an explicit norm-convergent Riemann integral. For (i), this is applied to $a_t$ and $a_s$ to deduce that there is a constant such that $\norm{f(D_t)-f(D_s)}\leq C \norm{a_t-a_s}$. If condition (ii) holds, the integrand in the Riemann integral lies in $\calK$  and, as $f(D)$ is invertible, this implies that $f(D + a)$ is Fredholm.
%
%
\hfill $\Box$

\vspace{.2cm}

If $T_0$ and $T_1$ are in $\Fsa$ and $T_0-T_1$ is compact, one can always consider the spectral flow along the straight-line path which will be denoted by
$$
\SF(T_0,T_1)
\;:=\;
 \SF(\{t T_1 + (1-t)T_0\})
 \;.
 $$
If $T_0$, $T_1$ and $T_2$ are in $\Fsa$ and both $T_0-T_1$ and $T_1-T_2$ are compact, then one has 
$$
\SF(T_0,T_1)
\;+\;
\SF(T_1,T_2)
\;=\;
\SF(T_0,T_2)
\;.
$$
%

\begin{definition}
Let $T \in \Nn$ be self-adjoint with a $\Tt$-finite (i.e. $\Tt$-trace-class) support projection. Then the signature of $T$ is defined by
$$
\Sig(T) 
\;:=\; 
\Tt(\sgn(T))
\;,
\qquad
\sgn(T)\;:=\;\chi(T > 0) \,-\, \chi(T < 0)
\;.
$$
\end{definition}

Let us note that the two summands $\chi(T > 0)$ and $\chi(T < 0)$ are separately trace-class. The following generalizes Sylvester's law of inertia. 

\begin{proposition}
Let $T \in \Nn$ be self-adjoint with a $\Tt$-finite support projection. Further let $A\in\Nn$ be invertible. Then 
$$
\Sig(A^*TA) 
\;=\; 
\Sig(T) 
\;.
$$
\end{proposition}

\noindent {\bf Proof.} 
Decomposing into positive and negative part $T= T_+ - T_-$, it is enough to prove the statement for $T\geq 0$. In that case, one has 
$$
\Sig(T)\;=\;\Tt(\Supp(T))\;, 
\qquad 
\Sig(A^*TA)\;=\;\Tt(\Supp(A^*TA))
\;,
$$ 
with the respective support projections. For $TA =  v\abs{TA}$ the unique polar decomposition with partial isometry $v$, one has $A^*T= \abs{TA}v^*=v^*v\abs{TA}v^*$ and thus 
$$
\Ran(v)\;=\;\overline{\Ran(TA)}\;=\;\overline{\Ran(T)}
\;,
\qquad
\Ran(v^*)\;=\;\overline{\Ran(A^*T)}=\overline{\Ran(A^*TA)}
\;.
$$
Since $T$ and $A^*TA$ are self-adjoint, their support projections are given by $\Supp(T)=v^*v$ and $\Supp(A^*TA)=vv^*$. Hence the claim follows from $\Tt(v^*v)=\Tt(vv^*)$.
\hfill $\Box$

\vspace{.2cm}

Certain spectral flows can be computed in terms of the signature:

\begin{proposition}
\label{prop:sf_as_signature}
Let $\{T_t\}$ be a norm-continuous path of self-adjoints in $\Nn$ such that the support projections satisfy $\Supp(T_t) \leq P$ and the range projections satisfy $\Ran(T_t) \leq P$ for all $t$ and a single $\Tt$-finite projection $P \in \Nn$. Then $\{F_t\}:= \{P T_t P + 1-P\}$ is a norm-continuous path in $\Fsa$ and 
\begin{equation} 
\label{eq:sig_formula}
\SF(\{F_t\}) 
\;=\; 
\tfrac{1}{2}\Big(\Sig(T_1) - \Sig(T_0)\Big) 
\,+\, 
\tfrac{1}{2}\Big(\Tt\big(P \Ker(T_1)\big)\,-\,\Tt\big(P \Ker(T_0)\big)
\Big)
\;.
\end{equation}
In particular, if the endpoints $T_0$ and $T_1$ are invertible elements of the algebra $P\Nn P$, then the second term vanishes.
\end{proposition}

\noindent {\bf Proof.} 
The Fredholm-property and continuity are obvious. Since $\pi(F_t) = 1$, the two point partition is sufficiently fine and hence the spectral flow is given by 
$$
\SF(\{F_t\}) 
\;=\; 
\ec(p_0,p_{1})
\;,
$$
with $p_t$ as in Definition~\ref{def-SF}. Since $P$ and $T_t$ commute, one has
$$
p_t 
\;=\; 
\chi(F_{t}>0)  
\;=\; 
\chi\big(PT_{t} P\oplus (1-P)>0\big) 
\;=\; 
P\chi(T_t>0)P \oplus (1-P)
\;,
$$
and hence
\begin{equation*}
\begin{split}
\Tt((1-p_0) \cap p_{1}) 
&
\;=\; 
\Tt\big((P(1-\chi(T_{0}>0))P)\cap(P\,\chi(T_{{1}}>0)P \big) 
\\
&\;=\; 
\Tt\big((P-P\,\chi(T_{0}>0)) \cap \chi(T_{{1}}>0) \big)
\\
&
\;=\;
\Tt\big(P  \,\chi(T_{{1}}>0)\big)\, -\, \Tt\big(P(\chi(T_{0}>0) \cap \chi(T_{{1}})>0) \big)
\;.
\end{split}
\end{equation*}
Switching $0$ and $1$ and taking the difference leads to
$$
\SF(\{F_t\}) 
\;=\; 
\Tt\big(P \, \chi(T_{1}>0)-P \, \chi( T_{0}>0)\big)
\;.
$$
Finally, noting that $P\,\chi(T_t>0) = \chi(T_t > 0)$ and thus 
$$ 
\Tt\big(P\,\Ker(T_t)\big) \,+\, \Tt\big(\chi(T_t > 0)\big) 
\;=\;  
\Tt\big(P\,\chi(T_t>0)\big) 
\;=\; 
\Tt(P) \,-\, 
\Tt\big(\chi(T_t < 0)\big)
\;,
$$ 
one obtains the formula \eqref{eq:sig_formula}.
\hfill $\Box$
 
\vspace{.2cm}

The signature also has an additional invariance property that is somewhat inconvenient to express in terms of the spectral flow: 

\begin{proposition}
\label{prop_signature_stability}
If $\{T_t\}_{t\in[0,1]}$ is a continuous path of self-adjoints all of which have compact support projections and such that for every $t\in[0,1]$ there is an open interval $\Delta_t$ around $0$ such that $\Delta_t \cap \sigma(T_t) \subset \{0\}$, then for all $t,t' \in [0,1]$
$$
\Sig(T_t)
\;=\;
\Sig(T_{t'})
\;.
$$
\end{proposition}

\noindent {\bf Proof.} 
As $[0,1]$ is compact, the spectra $\sigma(T_t) \setminus \{0\}$ have a common gap $\Delta$ and hence one can choose continuous functions $f,g$ such that
$$
\chi(T_t > 0) 
\;=\; 
f(T_t)
\;, 
\qquad 
\chi(T_t < 0) = g(T_t)\;,
 \qquad \forall \;t \in [0,1]\;.
 $$
Therefore the paths $t \mapsto \chi(T_t > 0)$ and $t \mapsto \chi(T_t < 0)$ are actually norm-continuous paths of projections. Since projections that are close in norm must be unitarily equivalent, this implies that the signature is constant along the path.
\hfill $\Box$

\section{Spectral localizer}
\label{sec-SpecLoc}

A semifinite spectral triple \cite{CareyEtAl,CGPR} (also called an unbounded or non-unital semifinite Fredholm module) is a tuple $(\calA, \Nn, D)$ consisting of a semifinite von Neumann algebra $\Nn$ with a semifinite faithful normal trace $\Tt$, a $*$-subalgebra $\calA \subset \Nn$ and a self-adjoint operator $D$ affiliated to $\Nn$ (namely, each spectral projection of $D$ lies in $\Nn$), which satisfy the following conditions:
\begin{enumerate}
\item[{\rm (i)}] For all $a \in \calA$ the commutator $[D,a]$ is densely defined and extends to a bounded operator (which is then an element of $\Nn$).
\item[{\rm (ii)}] For any $a \in \calA$ the product $a(1+D^2 )^{-\frac{1}{2}}$ is in $\calK$, {\it i.e.} is $\Tt$-compact.
\end{enumerate}
The triple is called {\it even} if there is a self-adjoint unitary $\gamma \in \Nn$ that commutes with all $a \in \calA$ and anticommutes with $D$, otherwise it is called {\it odd}. 
Moreover, the triple is called {\it Lipschitz-regular} if the commutators $[\abs{D},a]$ are also densely defined and extend to bounded operators for all $a\in \Aa$ \cite{CM} (notably this condition holds if the triple is $QC^1$ in the notation of {\rm \cite{CPRSI}}). It will also be assumed that $D$ is invertible. For the computation of index pairings this is no restriction, since one can apply the doubling construction from \cite{CGPR} if necessary: For any $\mu > 0$ the semifinite spectral triple $(\pi'(\calA), M_2(\Nn), D_\mu)$ with 
$$D_\mu 
\;=\; 
\begin{pmatrix}
D & \mu \\ - \mu & D
\end{pmatrix}
\;, 
\qquad 
\pi'(a) \;=\; \begin{pmatrix}
a & 0 \\ 0 & 0
\end{pmatrix}
$$
and in the even case with grading $\gamma'= \gamma \oplus (-\gamma)$ has an invertible Dirac operator and has the same index pairings as $(\Aa, \Nn, D)$. 
Furthermore, if $\Aa$ has no unit, let $\Aat\subset \Nn$ be its minimal unitization.

\vspace{.2cm}

For an odd spectral triple and a unitary $u \in \Aat$, one has the index pairing
$$
\langle [u],[D]\rangle 
\;:=\; 
\tInd_{(P\cdot P)}(P u P) 
\;,
$$
where $P = \chi(D>0)$. Under the condition $[D,u](1+D^2)^{-\frac{1}{2}} \in \calK$, the index pairing is connected to a spectral flow (see \cite{KNR} for the unital case and \cite{CGPR} for the nonunital one):
\begin{equation}
\label{eq-OddSpecFlow}
\langle [u],[D]\rangle 
\;= \;
\SF(u^* D u, D)
\;,
\end{equation}

\vspace{.2cm}

For an even spectral triple, the Dirac operator $D$ anti-commutes with a self-adjoint unitary $\gamma$ which is represented by the matrix $\diag(1,-1)$ in the grading provided by $\gamma$. As $D$ and all odd functions of $D$ are off-diagonal in this representation, there is an unbounded operator $D_0$ and a unitary $F$ such that
\begin{equation}
\label{eq-DiracEven}
D 
\;=\; 
\begin{pmatrix}
0 & D_0^* \\ D_0 & 0
\end{pmatrix}
\;, 
\qquad 
\sgn(D) 
\;=\; 
\begin{pmatrix}
0 & F^* \\ F & 0
\end{pmatrix}
\;.
\end{equation}
In the same sense, any element $a \in \Aat$ decomposes as $a=\diag(a_+,a_-)$ with respect to the grading of $\gamma$. 
For a projection $p \in \Aat$, the index pairing is defined by \cite{CPRS}
\begin{equation}
\label{eq-EvenIndPair}
\langle [p],[D]\rangle 
\; := \;
\tInd_{(p_+ \cdot p_-)}(p_+ F^* p_-) 
\;,
\end{equation}
where the properties of the spectral triple show that the skew-corner index is well-defined. It can be written as a spectral flow by
$$
\langle [p],[D]\rangle 
\; = \;
\ec(p_+, F^* p_- F)
\; = \;
\SF(1-2p_+, F^*(1-2p_-)F)
\;.
$$ 
If $p$ is given by $p = \chi(h < 0)$ for a self-adjoint invertible operator $h \in \Aat$, then continuously deforming $(1-2p)$ to $h$ shows 
\begin{equation}
\label{eq-SFEvenFormula}
\langle [p],[D]\rangle 
\; = \;
\SF(1-2p_+,F^*(1-2p_-)F)
\;=\; 
\SF(h_+, F^*h_-F)
\;.
\end{equation}
The straight-line paths (and the homotopy) are well-defined since $[a,\sgn(D)]$ is $\Tt$-compact for any $a\in \Aa$ and hence $F^* a_- - a_+ F^*$ is also $\Tt$-compact. We now define the spectral localizer:

\begin{definition}
Let $(\calA, \Nn, D)$ be a semifinite spectral triple and $\kappa \in \bbR_+$.
\begin{enumerate}
\item[{\rm (i)}]
If the triple is odd, assume that $a \in \calA^\sim$ is invertible and define the spectral localizer by
$$
L_{\kappa} 
\;:=\; 
\begin{pmatrix}
\kappa D &  a\\
a^*  & -\kappa D
\end{pmatrix} 
\;,
$$
as an operator affiliated to $M_2(\bbC)\otimes \Nn$. Setting $h= \begin{pmatrix}
0   &  a\\
a^*  &  0
\end{pmatrix}
$ and $D'=
\begin{pmatrix}
D   &  0\\
0  &  -D
\end{pmatrix}$,
the spectral localizer can also be written as $L_{\kappa} = h + \kappa D'$.

\item[{\rm (ii)}]
If $h \in \Aat$ is invertible and self-adjoint, the associated spectral localizer is defined by
$$
L_{\kappa} 
\;:=\;  \kappa D + h \gamma
\;=\;
\begin{pmatrix}
h_+ & \kappa D_0^*\\
\kappa D_0 & -h_-
\end{pmatrix}
\;,
$$
with the last expression again as a matrix w.r.t. the grading $\gamma$.

\end{enumerate}
Further set $P_\rho := \chi((D')^2 < \rho^2)$ in the odd case and $P_\rho := \chi(D^2 < \rho^2)$ in the even case. Then in both cases  the reduced spectral localizer is defined by
$$
L_{\kappa,\rho}
\;:=\; 
P_\rho L_\kappa P_\rho
\;.
$$
\end{definition}

The two cases of odd and even pairings are similar: there is a self-adjoint unitary that anti-commutes with $h$ and commutes with $D'$ in the odd case, respectively that anti-commutes with $D$ and commutes with $h$ in the even case. Both of the associated  pairings can now be read off the spectral localizer, as shows the main result of the paper:

\begin{theorem}
\label{th:main}
Let $(\calA, \Nn, D)$ be a semifinite spectral triple. In the even case, the triple is assumed to be Lipschitz-regular and $h\in \Aat$ to be invertible. In the odd case, let $h= \begin{pmatrix} 0   &  a\\ a^*  &  0 \end{pmatrix}$ with $a \in \Aat$ an invertible satisfying $[D,a] \in \Aa$. Further let $g=\norm{h^{-1}}^{-1}$ be the size of the spectral gap of $h$. If $\kappa > 0$ is chosen so small that
\begin{equation}
\label{eq:kapparho1} 
\kappa 
\;\leq \;
\frac{g^3}{12 \norm{[D,h]}\, \norm{h}}
\;,
\end{equation}
and $\rho$ so large that 
\begin{equation}
\label{eq:kapparho2}
\rho\; > \;\frac{2g}{\kappa}
\;,
\end{equation}
then the spectral localizer $L_{\kappa,\rho}$ as defined above is invertible and satisfies
$$
\langle [a \abs{a}^{-1}],[D]\rangle 
\;=\; 
\tfrac{1}{2} \;\Sig(L_{\kappa,\rho})
$$
in the odd case and
$$
\langle [\chi(h<0)],[D]\rangle 
\;=\; 
-\;\tfrac{1}{2} \;\Sig(L_{\kappa,\rho})
$$
in the even case.
\end{theorem}

Let us note that upon scaling $h\mapsto \lambda h$ with $\lambda>0$, one has $g\mapsto \lambda g$ so that  by \eqref{eq:kapparho1} also $\kappa\mapsto\lambda\kappa$ and hence the condition \eqref{eq:kapparho2} on $\rho$ remains unchanged. In this sense, the bounds are natural.

\section{Constancy of the signature}
\label{sec-Constancy}

For a given $\rho>0$, let us introduce a smooth cut-off function $G \in C^\infty(\bbR)$ such that 
$$
G_\rho(x) 
\;=\; 
\begin{cases} 1\;, \quad \text{for }\abs{x} \leq \frac{\rho}{2} \;,
\\ 0
\;, \quad \text{for }\abs{x} > \rho
\;,
\end{cases}
$$
and whose Fourier transform satisfies $\norm{\calF (G_\rho')}_1 \leq 8 \rho^{-1}$. Such a function is constructed in \cite[Section 3]{LS} and it is shown that
%
$$
\norm{[G_\rho(D), a]} \;\leq\; 
\norm{\mathcal{F}(G'_\rho)}_1 \norm{[D,a]}
\;\leq\; 
\frac{8}{\rho} \norm{[D,a]}
\;.
$$ 

\begin{lemma}
\label{lem:inv_const}
A pair $(\kappa,\rho)$ is called admissible if it satisfies the inequalities \eqref{eq:kapparho1} and \eqref{eq:kapparho2}.

\vspace{-.3cm}

\begin{enumerate}
\item[{\rm (i)}] If $(\kappa,\rho)$ is admissible, then $(L_{\kappa,\rho})^2 \;>\; \frac{g^2}{4}\, P_\rho$.

\vspace{-.2cm}

\item[{\rm (ii)}] If $(\kappa,\rho)$ and $(\kappa',\rho')$ are admissible, then $\Sig(L_{\kappa,\rho})=\Sig(L_{\kappa',\rho'})$.
\end{enumerate}
\end{lemma}

\vspace{-.1cm}

\noindent {\bf Proof.}
Let us focus on the proof in the odd case, following \cite[Section 2]{LS3}. The even case can be dealt with by similar means as in \cite{LS2} (it is enough to replace $H\otimes \sigma_0$ and $H \otimes \Gamma$ appearing in the proof of Theorem 3 of \cite{LS2} with $h$ and $h \gamma$ respectively). For $\rho \leq \rho'$, let us introduce the path
\begin{equation}
\label{eq:stability_path1}
\lambda\in[0,1]\;\mapsto\;
L_{\kappa, \rho, \rho'}(\lambda) \,:=\, \kappa P_{\rho'} D' P_{\rho'} \,+\, P_{\rho'} G_{\lambda,\rho} h G_{\lambda,\rho} P_{\rho'}
\;,
\end{equation}
with $G_{\lambda,\rho} = (1-\lambda) + \lambda G_\rho(D')$. By literally the same computation as in \cite{LS3}, one obtains
$$
(L_{\kappa, \rho, \rho'}(\lambda))^2 + (1-P_{\rho'}) 
\;>\; 0
\;,
\qquad
(L_{\kappa, \rho, \rho'}(0))^2 
\;>\; 
\frac{g^2}{4} \,P_{\rho'}
\;,
$$
provided that $(\kappa,\rho)$ is admissible. In particular, $(L_{\kappa, \rho})^2 = (L_{\kappa, \rho, \rho}(0))^2 >  \frac{g^2}{4} P_{\rho'}$. Next let us show that $\Sig(L_{\kappa,\rho}) = \Sig(L_{\kappa',\rho'})$ for any admissible pairs $(\kappa,\rho)$ and $(\kappa',\rho')$ with  $\rho \leq \rho'$. As $L_{\lambda\kappa'+(1-\lambda)\kappa,\rho'}$ is gapped around $0$ for any $\lambda$, the signature remains unchanged by Proposition~\ref{prop_signature_stability}, hence one may assume $\kappa=\kappa'$. Since the path \eqref{eq:stability_path1} is continuous and also satisfies the conditions of Proposition \ref{prop_signature_stability}, it is sufficient to prove $\Sig(L_{\kappa, \rho, \rho}(1)) = \Sig(L_{\kappa, \rho, \rho'}(1))$. As $P_{\rho'}G_\rho=P_\rho G_\rho$, one has
$$
L_{\kappa,\rho,\rho'}(1)
\;=\; 
\kappa P_\rho D' P_\rho + P_{\rho'} G_{\rho} h G_{\rho} P_{\rho'} 
\;=\; 
L_{\kappa,\rho,\rho}(1) + \kappa (P_{\rho'}-P_{\rho})D' (P_{\rho'}-P_{\rho})
\;,
$$
and, moreover, the last sum is direct. Hence
$$
\Sig(L_{\kappa,\rho,\rho'}(1))
\;=\;
\Sig(L_{\kappa,\rho,\rho}(1))\,+\,\Sig((P_{\rho'}-P_{\rho})D' (P_{\rho'}-P_{\rho})) 
\;=\; 
\Sig(L_{\kappa,\rho,\rho}(1))
\;,
$$
where the second equality holds obviously due to the definition of $D'$.
\hfill $\Box$

\vspace{.2cm}

For $a$ affiliated to $\Nn$ or matrices over $\Nn$, let us write $a_\rho = P_\rho a P_\rho$ and $a_{\rho^c}=(1-P_\rho)a(1-P_\rho)$. Then $D= D_\rho + D_{\rho^c}$ and, in the odd and even case respectively,
\begin{align*}
L_{\kappa} 
\;=\; L_{\kappa,\rho}\oplus L_{\kappa,\rho^c} \,+\, P_\rho h P_{\rho^c} \,+\,
P_{\rho^c} h P_\rho\;,
\quad
L_{\kappa} 
\;=\; L_{\kappa,\rho}\oplus L_{\kappa,\rho^c} \,+\, P_\rho (h\gamma) P_{\rho^c} \,+\, P_{\rho^c}\,(h\gamma) P_{\rho}
\;.
\end{align*}

\begin{lemma}
\label{lem:localizer_decoupling}
If $(\kappa, \rho)$ is admissible and $\rho$ is large enough, then $L_{\kappa,\rho^c}$ is invertible in $P_{\rho^c}\Nn P_{\rho^c}$ and $\SF(L_{\kappa},L_{\kappa,\rho}\oplus L_{\kappa,\rho^c}) =0$.
\end{lemma}

\noindent {\bf Proof.}
We will show that for $\rho$ large enough the term of $L_\kappa$ that is off-diagonal with respect to the decomposition $\one = P_\rho \oplus P_{\rho^c}$ can be shrunk to zero with a linear path $t\in[0,1]\mapsto L_\kappa(t)$ in the invertible operators given by
$$
L_{\kappa}(t)
\;=\; 
L_{\kappa,\rho}\oplus L_{\kappa,\rho^c} 
\,+\, 
t(P_\rho L_\kappa P_{\rho^c} +
P_{\rho^c}\,L_\kappa P_\rho)
 \;.
$$
Let us focus on the odd case as the even case follows by essentially the same argument. The diagonal part is invertible because $\abs{L_{\kappa,\rho}} > \frac{g}{2}P_\rho$ by Lemma~\ref{lem:inv_const} and 
\begin{align} 
(L_{\kappa,\rho^c})^2 
&
\;=\; 
\begin{pmatrix}
\kappa^2 D_{0,\rho^c}^2 + a_{\rho^c} a_{\rho^c}^* & \kappa [D_{0,\rho^c}, a_{\rho^c}] \\ 
\kappa [D_{0,\rho^c}, a_{\rho^c}]^* & \kappa^2 D_{0,\rho^c}^2 + a_{\rho^c}^* a_{\rho^c}
\end{pmatrix} 
\nonumber
\\ &
\;\geq \;
\kappa^2 \rho^2 P_{\rho^c} \one_2 
\,+\, 
\kappa
\begin{pmatrix}
0 & [D_{0,\rho^c}, a_{\rho^c}] \\ 
[D_{0,\rho^c}, a_{\rho^c}]^* & 0
\end{pmatrix} 
\nonumber
\\
&
\;\geq \;
(\kappa^2\rho^2 - \kappa \norm{[D_0,a]})P_{\rho^c}\one_2 
\;\geq \;
\tfrac{1}{2}\,\kappa^2\rho^2 P_{\rho^c}\one_2
\;,
\label{eq-RhocBound}
\end{align}
due to \eqref{eq:kapparho1} and for $\rho$ sufficiently large.
The inverse is again diagonal in the decomposition $\one = P_\rho \oplus P_{\rho^c}$ and given by
$$
\abs{L_{\kappa,\rho}\oplus L_{\kappa,\rho^c}}^{-\frac{1}{2}} 
\;=\; 
P_\rho\abs{L_{\kappa,\rho}}^{-\frac{1}{2}} 
\oplus 
P_{\rho^c}\abs{L_{\kappa,\rho^c}}^{-\frac{1}{2}}
\;,
$$
such that $L_{\kappa}(t)$ is equal to
$$
\abs{L_{\kappa,\rho}\oplus L_{\kappa,\rho^c}}^{\frac{1}{2}} 
\Big(
S + 
t 
\abs{L_{\kappa,\rho}}^{-\frac{1}{2}} P_\rho h  P_{\rho^c} \abs{L_{\kappa,\rho^c}}^{-\frac{1}{2}} +  
t
\abs{L_{\kappa,\rho^c}}^{-\frac{1}{2}} P_{\rho^c}h P_\rho
\abs{L_{\kappa,\rho}}^{-\frac{1}{2}}
\Big)
\abs{L_{\kappa,\rho}\oplus L_{\kappa,\rho^c}}^{\frac{1}{2}} 
\;,
$$
with $S$ being the unitary from the polar decomposition of $L_{\kappa,\rho}\oplus L_{\kappa,\rho^c}$. By a Neumann series argument, the term in brackets and thus $L_\kappa(t)$ is invertible for $t \leq 1$ if the norm of the off-diagonal component is less than $1$. Since
$$
\norm{\abs{L_{\kappa,\rho}}^{-\frac{1}{2}} P_\rho h  P_{\rho^c} \abs{L_{\kappa,\rho^c}}^{-\frac{1}{2}}}
\; \leq \;
\frac{ 2^{\frac{3}{4}} \norm{h}}{g^{\frac{1}{2}} (\rho^2 \kappa^2)^{\frac{1}{4}}} 
\;,
$$
this is the case for $\rho$ large enough. 
\hfill $\Box$

\section{Proof of the odd index formula}
\label{sec-ProofOdd}

This section provides the proof of the odd case of Theorem~\ref{th:main}. Thus let us consider an odd spectral triple and  suppose that $(\kappa, \rho)$ is admissible. By Lemma~\ref{lem:inv_const} (ii) it is enough to prove the signature formula for some admissible $(\kappa,\rho)$ and we assume $\rho$ to be large enough for Lemma~\ref{lem:localizer_decoupling} to hold. As $a$ is invertible, its polar decomposition  $u= a\abs{a}^{-1}$ can be written using the Riesz projection $\frac{1}{2}\begin{pmatrix}
\one & u \\
u^* & \one
\end{pmatrix} = \chi(h>0) = \frac{1}{2\pi \imath}\int_{\mathcal{C}} (h-z)^{-1}\difd{z}
$ for a positively oriented contour $\mathcal{C}$ around $\sigma(\abs{a})\cap \bbR^+$. Hence $u$ is an element of the unitization of the norm-completion $\overline{\Aa}^\sim$. 
As $[D,a] \in \Aa$ by assumption, the resolvent identity $[D,(h-z)^{-1}]=- (h-z)^{-1}[D, h](h-z)^{-1}$ shows that $[D,u]$ densely defines an element of $\overline{\Aa}$, hence $[D,u](1+D^2)^{-\frac{1}{2}} \in \calK$. The spectral flow formula \eqref{eq-OddSpecFlow} thus gives
\begin{equation*}
\begin{split}
\langle [u],[D]\rangle  
&
\;=\; 
\SF\left(\begin{pmatrix}
u^* & 0 \\ 0 & 1
\end{pmatrix}
\begin{pmatrix}
\kappa D & 0 \\ 0 & -\kappa D
\end{pmatrix}
\begin{pmatrix}
u & 0\\ 0 & 1
\end{pmatrix}
,\begin{pmatrix}
\kappa D & 0 \\ 0 & -\kappa D
\end{pmatrix}
\right)\\
&
\;=\; 
\SF\left(\begin{pmatrix}
\kappa D & 0 \\ 0 & -\kappa D
\end{pmatrix},\begin{pmatrix}
u & 0 \\ 0 & 1
\end{pmatrix}
\begin{pmatrix}
\kappa D & 0 \\ 0 & -\kappa D
\end{pmatrix}
\begin{pmatrix}
u^* & 0\\ 0 & 1
\end{pmatrix}
\right)
\;.
\end{split}
\end{equation*}
%
%
Noting that the path is in the invertibles except at the left endpoint, which has by assumption a trivial kernel, one must have 
$$
\SF\left(\begin{pmatrix}
\kappa D & 0 \\ 0 & -\kappa D
\end{pmatrix},\begin{pmatrix}
\kappa D & 1 \\ 1 & -\kappa D
\end{pmatrix}\right)
\;=\;
0
\;,
$$
and hence also
$$
\SF\left(
\begin{pmatrix}
u & 0 \\ 0 & 1
\end{pmatrix}
\begin{pmatrix}
\kappa D & 0 \\ 0 & -\kappa D
\end{pmatrix}
\begin{pmatrix}
u^* & 0 \\ 0 & 1
\end{pmatrix},
\begin{pmatrix}
u & 0 \\ 0 & 1
\end{pmatrix}
\begin{pmatrix}
\kappa D & 1 \\ 1 & -\kappa D
\end{pmatrix}
\begin{pmatrix}
u^* & 0 \\ 0 & 1
\end{pmatrix}
\right)
\;=\;
0
\;.
$$
Using the concatenation property of the spectral flow and deforming the resulting path again to a straight-line path with a homotopy in the space of Fredholm operators implies
\begin{align*}
\langle [u],[D]\rangle  
&
\;=\; 
\SF\left(\begin{pmatrix}
\kappa D & 0 \\ 0 & -\kappa D
\end{pmatrix},\begin{pmatrix}
u & 0 \\ 0 & 1
\end{pmatrix}
\begin{pmatrix}
\kappa D & 1 \\ 1 & -\kappa D
\end{pmatrix}
\begin{pmatrix}
u^* & 0\\ 0 & 1
\end{pmatrix}
\right)
\\
&
\;=\; 
\SF\left(\begin{pmatrix}
\kappa D & 0 \\ 0 & -\kappa D
\end{pmatrix},
\begin{pmatrix}
\kappa\, uDu^* & u \\ u^* & -\kappa D
\end{pmatrix}
\right)
\;.
\end{align*}
Now $\kappa uDu^*=\kappa D+\kappa u[D,u^*]$ and $\kappa u[D,u^*]$ is a bounded summand that for $\kappa$ sufficiently small does not alter the invertibility of the spectral localizer of $u$ so that
$$
\SF\left(
\begin{pmatrix}
\kappa uDu^* & u \\ u^* & -\kappa D
\end{pmatrix},
\begin{pmatrix}
\kappa D & u \\ u^* & -\kappa D
\end{pmatrix}
\right)
\;=\;0
\;.
$$
Therefore again by concatenation and deforming the paths one obtains
\begin{equation*}
\langle [u],[D]\rangle  
\;=\; 
\SF\left(\begin{pmatrix}
\kappa D & 0 \\ 0 & -\kappa D
\end{pmatrix},
\begin{pmatrix}
\kappa D & u \\ u^* & -\kappa D
\end{pmatrix}
\right)
\;.
\end{equation*}
To see that this and similar straight-line paths are well-defined in the non-unital case let us note that it becomes a continuous path under the bounded transform by Proposition~\ref{eq-DiracPerturb}(i) and is pointwise Fredholm due to the decomposition 
$$\begin{pmatrix}
\kappa D & u \\ u^* & -\kappa D
\end{pmatrix} 
\;=\; 
\begin{pmatrix}
\kappa D & s(u) \\ s(u)^* & -\kappa D
\end{pmatrix} 
\;+\; 
\begin{pmatrix}
0 & u-s(u) \\ u^*-s(u)^* & 0
\end{pmatrix}
\;,
$$ 
where $s(u)$ is the scalar part of $u\in \overline{\Aa}^\sim$. The first term squares to $(\kappa^2 D^2 + \abs{s(u)}^2)\one_2$, hence the second term is a bounded perturbation of an invertible self-adjoint operator that satisfies the conditions of Proposition~\ref{eq-DiracPerturb}(ii) due to $u-s(u)\in\overline{\Aa}$. Furthermore, one can connect $u$ to $a$ by the path $t\in[0,1]\mapsto a(t +(1-t)\abs{a}^{-1})$ leading to a path of invertible spectral localizers. Hence again by concatenation and deformation of the straight-line paths
\begin{equation*}
\langle [u],[D]\rangle  
\;=\; 
\SF\left(\begin{pmatrix}
\kappa D & 0 \\ 0 & -\kappa D
\end{pmatrix},
\begin{pmatrix}
\kappa D & a \\ a^* & -\kappa D
\end{pmatrix}
\right)
\;=\;
\SF\left(\begin{pmatrix}
\kappa D & 0 \\ 0 & -\kappa D
\end{pmatrix},
L_\kappa
\right)
\;.
\end{equation*}
Now by Lemma~\ref{lem:localizer_decoupling} one can decouple $L_\kappa$ to $L_{\kappa,\rho}\oplus L_{\kappa,{\rho^c}}$ so that, once again by concatenation and deforming the linear paths,
$$
\langle [u],[D]\rangle  
\;=\;
\SF\left(\begin{pmatrix}
\kappa D & 0 \\ 0 & -\kappa D
\end{pmatrix},
L_{\kappa,\rho}\oplus L_{\kappa,{\rho^c}}
\right)
\;.
$$
Finally also the first entry is diagonal in the decomposition $P_\rho \oplus P_{\rho^c}$ so that
\begin{align*}
\langle [u],[D]\rangle  
&
\;=\;
\SF\left(
\begin{pmatrix}
\kappa D_\rho & 0 \\ 0 & -\kappa D_\rho
\end{pmatrix}
\oplus 
\begin{pmatrix}
\kappa D_{\rho^c} & 0 \\ 0 & -\kappa D_{\rho^c}
\end{pmatrix}
, L_{\kappa,\rho}\oplus L_{\kappa,{\rho^c}}
\right)
\\
&
\;=\; 
\SF\left(\begin{pmatrix}
\kappa D_\rho & 0 \\ 0 & -\kappa D_\rho
\end{pmatrix}
, L_{\kappa,\rho}
\right)
\; + \SF\left(\begin{pmatrix}
\kappa D_{\rho^c} & 0 \\ 0 & -\kappa D_{\rho^c}
\end{pmatrix}
, L_{\kappa,{\rho^c}}
\right)
\end{align*}
because one can use the homomorphism property of Proposition~\ref{prop-SFprop}(iv) and the fact that the spectral flow on $P_{\rho^c}$ vanishes since the path is in the invertibles due to \eqref{eq-RhocBound}. Finally, the signature formula Proposition~\ref{prop:sf_as_signature} completes the proof  of the odd case of Theorem~\ref{th:main} by noting that the signature of the left endpoint vanishes. 

\section{Proof of the even index formula}
\label{sec-ProofEven}

In the even case, the spectral triple $(\Aa,\Nn,D)$ is assumed to be Lipschitz regularity so that not only $[D, a]$ is bounded for every $a\in \Aa$, but $[\abs{D},a]$ as well. Hence the (in general inequivalent) representation of $\Aa$ in $\Nn$ given by 
$$
a \in \Aa 
\;\mapsto\; 
\pi^+(a) 
\;=\;
\begin{pmatrix}
a_+ &  0\\
0 & F a_+ F^*
\end{pmatrix}
\;
$$
also defines an even spectral triple $(\pi^+(\calA), \Nn, D)$. However, writing out \eqref{eq-EvenIndPair} one finds that its index pairing vanishes. The spectral localizer with respect to this triple is still useful and is denoted by
$$
L_{\kappa}^+
\;:=\; 
\begin{pmatrix}
h_+ & \kappa D_0^*\\
\kappa D_0 & -F h_+ F^*
\end{pmatrix}
\;,
$$
and analogously $L_{\kappa,\rho}^+ = P_\rho L_\kappa^+ P_\rho$.   Note that in a spectral triple $[a,\sgn(D)]$ is $\Tt$-compact for all $a\in \Aa$ and hence also $a -\pi^+(a) \in \Kk$, so {\it e.g.} the spectral flow from $L_\kappa$ to $L^+_\kappa$ is well-defined.

\vspace{.1cm}

The starting point for the proof of the even signature formula is the spectral flow formula $\langle [p],[D]\rangle = \SF(F h_+ F^*, h_-)$ given in \eqref{eq-SFEvenFormula}. The additivity of the spectral flow leads to
$$
\langle [p],[D]\rangle
\;=\;
\SF\left(
\begin{pmatrix}
h_+ & 0\\
0    & -h_-
\end{pmatrix}
,
\begin{pmatrix}
h_+ & 0\\
0    & -F h_+ F^*
\end{pmatrix}
\,
\right)
\;=\; 
\SF\left(
h \gamma
,
\pi^+(h) \gamma
\,
\right)
\;.
$$
Again by Lemma \ref{lem:inv_const}(ii) it is enough to prove the signature formula for some admissible $(\kappa,\rho)$, so one can assume that $\kappa$ is as small and $\rho$ as large as necessary.

\begin{lemma}
\label{lem:tech1}
For $\kappa$ small enough, one has $\SF(h \gamma, L_\kappa) = 0$ and $\SF(\pi^+(h) \gamma, L^+_\kappa) = 0$.
\end{lemma}

\noindent {\bf Proof.}
Considering that
$$
\big((1-t) (h  \gamma) + tL_\kappa\big)^2 
\;=\; h^2 + t^2 \kappa^2 D^2 + t \kappa [h,D] 
\;\geq\; 
(g^2 - \kappa \norm{[D,h]})\one
\;,
$$
the straight-line path connecting $h \gamma$ and $L_\kappa$ is invertible for $\kappa \norm{[D,h]}<g^2$. The same holds for the other path since $(\pi^+(\calA), \Nn, D)$ is also a spectral triple.
\hfill $\Box$
 
\vspace{.1cm} 

Since the intermediate paths have compact differences, we can concatenate and then deform back to a straight-line path, hence
$\langle [p],[D]\rangle=\SF\left(L_\kappa, L^+_\kappa\right)$.
%
%
According to Lemma~\ref{lem:localizer_decoupling} the endpoints decouple, namely 
$$
\SF(L_\kappa, L_{\kappa,\rho} \oplus L_{\kappa, \rho^c}) 
\;=\; 
0 
\;=\; 
\SF(L^+_\kappa, L^+_{\kappa,\rho} \oplus L^+_{\kappa, \rho^c})
$$ 
for sufficiently large $\rho$. The paths that shrink the off-diagonal parts of the localizers to zero again have compact differences, so the spectral flow can be decomposed into two summands using the additivity. The contribution on $P_\rho$ can be expressed in terms of the signature due to Proposition~\ref{prop:sf_as_signature}:
\begin{align*}
\SF(L_\kappa, L^+_\kappa)
\,=\,
\SF(L_{\kappa,\rho}, L^+_{\kappa, \rho}) \;+\; \SF(L_{\kappa,\rho^c},  L^+_{\kappa, \rho^c}) 
\,=\, 
\tfrac{1}{2}\big(\Sig(L^+_{\kappa,\rho})\,-\, \Sig(L_{\kappa,\rho}))  
\,+\, 
\SF(L_{\kappa,\rho^c},  L^+_{\kappa, \rho^c})
\,.
\end{align*}
The signature $\Sig(L^+_{\kappa,\rho})$ vanishes since $
L^+_{\kappa,\rho} 
=
\begin{pmatrix}
0 & F^*\\ -F & 0
\end{pmatrix} (-L^+_{\kappa,\rho}) \begin{pmatrix}
0 & -F^*\\ F & 0
\end{pmatrix}
$ is unitarily equivalent to its negative. It only remains to show that the last summand also vanishes:

\begin{lemma}
\label{lem:tech2}
For $\kappa \rho$ large enough, one has $\SF(L^+_{\kappa,\rho^c},  L_{\kappa, \rho^c})=0.$
\end{lemma}

\noindent{\bf Proof.} Again let us consider the square
\begin{align*}
&
((1-t)L^+_{\kappa,\rho^c} + t L_{\kappa,\rho^c})^2 
\\
&
\;\;\;\;\;= \;\kappa^2 D_{\rho^c}^2 + t\kappa[h_{\rho^c} , D_{\rho^c}] + (1-t)\kappa [\pi^+(h)_{\rho^c}, D_{\rho^c}] 
\\ 
&
\;\;\;\;\;\;\;\;\;\;\;\;
+\; \begin{pmatrix}
(h_+)^2_{\rho^c} & 0\\
0 & t^2(h_-)^2_{\rho^c} + (1-t)^2 (F h_+ F^*)^2_{\rho^c} + t(1-t) [(h_-)_{\rho^c},(F h_+ F^*)_{\rho^c}]
\end{pmatrix}
\\
&\;\;\;\;\;\geq \;(\kappa^2 \rho^2 - \kappa \norm{[D,h]} - \kappa \norm{[D,\pi^+(h)]} - \norm{h}^2)\,\one_{\rho^c}
\end{align*}
where $h_{\rho^c}=P_{\rho^c} h P_{\rho^c}$ and $(h_{\pm})_{\rho^c} = P^{\pm}_{\rho^c} h_{\pm} P^{\pm}_{\rho^c}$ with $P_{\rho^c} = \diag(P^+_{\rho^c} , P^-_{\rho^c})$. This shows that the path lies within the invertibles for large $\kappa\rho$.
\hfill $\Box$

\section{Application to topological insulators}
\label{sec-app}

If the semifinite von Neumann algebra $(\Nn,\Tt)$ is of type $\text{I}_\infty$, namely given by a pair $(\Bb(\Hh),\Tr)$, then the finite volume spectral localizer $L_{\kappa,\rho}$ is a finite-dimensional matrix and it is immediately possible to use it for numerical computation of the index pairing based on Theorem~\ref{th:main}, see \cite{Lor,Lor2,LSS}. In the setting of a type $\text{II}_\infty$-von Neumann algebra, the spectral localizer $L_{\kappa,\rho}$ is in general an operator of infinite rank, but its signature may still be well approximated by finite-dimensional quantities that are accessible to numerical computations. Here we sketch how this can be done for a large class of Schr\"odinger-type operators describing topological insulators. A typical example for an observable algebra is the disordered non-commutative torus $\Aa = C(\Omega)\rtimes_\xi \bbZ^d$, constructed from an invariant ergodic probability space $(\Omega,\ZM^d,\mathbb{P})$ describing homogeneous disorder and a twist $\xi$ provided by the magnetic field \cite{PSbook}. In the representation on $\ell^2(\bbZ^d)$, one considers for $n$ lattice directions $e_1,\ldots,e_n$ the Dirac operator $D = \sum_{k=1}^n \sigma_k \otimes (\frac{1}{2}\one + X_k)$ (shifted to ensure invertibility) with $\sigma_k$ a representation of the complex Clifford algebra of $n$ generators and $X_1,\ldots,X_n$ the unbounded position operators. For $n<d$, the non-integer-valued index pairing with $D$ is a so-called weak Chern number and is localized by a spectral triple $(\calA, \Nn, D)$ ({\it cf.} \cite{BP,BSB}), where $\Nn$ is the von Neumann algebra generated by $\Aa$ and bounded functions of $X_1,\ldots,X_n$ equipped with a trace $\Tt$ that can be interpreted as an average trace per volume. A self-adjoint invertible $h \in M_N(\Aa_d)$, assumed to take the form $h=\begin{pmatrix}
0 & a^*\\ a & 0
\end{pmatrix}$ for $n$ odd, describes a random family $(h_\omega)_{\omega\in \Omega}$ of Hamiltonians on $\ell^2(\bbZ^d,\CM^N)$ modeling a topological insulator and the index pairings $\langle [\chi(h \leq 0)], [D]\rangle$, respectively $\langle[a\abs{a}^{-1}],[D]\rangle$, are related to linear response coefficients and the appearance of topologically protected boundary states \cite{PSbook}. 

\vspace{.1cm}

The spectral localizer $L_{\kappa,\rho}=(L_{\kappa,\rho,\omega})_{\omega\in \Omega}$ can be considered a random family of Schr\"odinger-type Hamiltonians acting on the Hilbert space $\ell^2(B^n_\rho \times \bbZ^{d-n}, \CM^{N'})$ with $B_\rho^n = \{x \in \bbZ^n: \sum_{i=1}^n (x_i+\frac{1}{2})^2 \leq \rho^2\}$ and $N'$ a large enough fiber dimension to accommodate both the Dirac operator $D$ and $h\oplus (-h)$ in the even case, respectively $D'$ and $h$ in the odd case. It can be interpreted as a perturbation of the restriction of $h$ to the cylinder $B_\rho^n \times \bbZ^{d-n}$ by an additional potential $\kappa D_\rho$. For numerical computations, one further truncates to operators $L^{(V_{\ell})}_{\kappa,\rho,\omega}$ acting on the finite-dimensional space $\ell^2(B_\rho^n \times V_\ell, \CM^{N'})$ with $V_\ell$ being a cube with sides $\ell$ and supplied with {\it e.g.} periodic or Dirichlet boundary conditions. For $h$ satisfying the usual smoothness conditions and $f \in C_0(\bbR)$, one can show that almost surely with respect to $\PM$
$$
\Tt(f(L_{\kappa,\rho})) 
\;\stackrel{\text{a.s.}}{=}\; \;
\lim_{\ell\to \infty}\, \frac{\abs{B_\rho^n}}{\abs{V_\ell}} \;\Tr(f(L^{(V_{\ell})}_{\kappa,\rho,\omega}))
\;.
$$
Since $L_{\kappa,\rho}$ has a spectral gap, one can replace $\Sig(L_{\kappa,\rho})=\Tt(\sgn(L_{\kappa,\rho}))=\Tt(f(L_{\kappa,\rho}))$ for a suitable continuous function $f$ and hence the signature can be approximated using only finite-dimensional algebra. When choosing restrictions with periodic boundary conditions, one can adapt methods from \cite{Prodan2017} to show that the approximations $L^{(V_{\ell})}_{\kappa,\rho,\omega}$ have a uniform spectral gap and
$$
\Sig(L_{\kappa,\rho}) 
\,\stackrel{\text{a.s.}}{=}\;
\lim_{\ell\to \infty} \frac{\abs{B_\rho^n}}{\abs{V_\ell}} \Big[ \#(\text{positive eigenvalues of }L^{(V_{\ell})}_{\kappa,\rho,\omega})-\#(\text{negative eigenvalues of }L^{(V_{\ell})}_{\kappa,\rho,\omega}) \Big]
,
$$ 
where the deterministic component $\big|\Sig(L_{\kappa,\rho})-|B_\rho^n|\,|V_\ell|^{-1}\, \mathbb{E}\, \Sig(L^{(V_{\ell})}_{\kappa,\rho,\omega})\big|$ of the finite volume error is exponentially small in $\ell$ for typical Hamiltonians. We expect that the method described here generalizes to compute weak invariants of other aperiodic quantum systems which are {\it e.g.} described by point patterns \cite{BP}.


\end{document}